\documentclass[fleqn]{article}

\def\GeV{{\rm\ GeV}}
\def\ve{\varepsilon}
\def\CG{{\cal G}}
\def\CM{{\cal M}}
\def\be{\begin{equation}}
\def\ee{\end{equation}}
\def\bea{\begin{eqnarray}}
\def\eea{\end{eqnarray}}
\def\Re{\mathop{\rm Re}\nolimits}
\def\Im{\mathop{\rm Im}\nolimits}
\def\tg{\mathop{\rm tg}\nolimits}

\begin{document}

\title{{\tt TPEcalc}: a program for calculation\\
 of two-photon exchange amplitudes}

\author{Dmitry~Borisyuk and Alexander~Kobushkin\\[5mm]
\it Bogolyubov Institute for Theoretical Physics,\\
\it 14-B Metrologicheskaya street, Kiev 03680, Ukraine
}


\date{}

\maketitle

\begin{abstract}
  {\tt TPEcalc} is a C++ program for calculation of two-photon exchange (TPE)
  amplitudes in elastic electron-hadron scattering,
  based on the dispersion method.
  It is a command-line tool which accepts kinematical parameters
  ($Q^2$ and $\ve$) as input and returns TPE amplitudes.
  It can do calculations for proton, neutron and pion targets.
  Any spin-$1/2^+$ or spin-$0^-$ target is supported,
  if the user supplies all necessary form factor parameterizations.
  This paper describes how to use {\tt TPEcalc}
  and outlines underlying theory.
  The program source code can be downloaded from
  {\tt http://tpe.bitp.kiev.ua/}.
\end{abstract}

\section{Introduction}
During last decade, two-photon exchange (TPE) has drawn much attention
of both theorists and experimentalists.
It was shown that TPE corrections remove the discrepancy
between Rosenbluth and polarization methods
in proton form factors measurements.
The TPE corrections have been applied to low-$Q^2$ \cite{BlundenSick}
and high-$Q^2$ \cite{BlundenRes,Arrington} elastic $ep$ scattering.
Many experimental searches for TPE effects are proposed
(see e.g. \cite{TPEexp} and references wherein)
and some are already finished \cite{TPEfin}.

The theoretical evaluation of TPE effects was done by several authors
\cite{BlundenRes,GPD,ourDisp}.
However, at present there is no publicly available computer program
for the calculation of TPE corrections,
though the need for it clearly exists.
Herewith we present the program {\tt TPEcalc}, which is capable of 
calculating TPE amplitudes for the elastic $ep$, $en$ and $e\pi$ scattering,
with a variety of intermediate states (ISs).
The calculation is based on the dispersion method \cite{ourDisp},
see also \cite{ourPi}.
It closely follows the procedure, described in Sec.~III of Ref.~\cite{ourDisp},
thus there is no detailed discussion here.

The program is written in C++. 
It is a command-line tool which accepts
kinematical parameters ($Q^2$ and $\ve$) as input
and returns TPE amplitudes.
It has such options as
customizing output format, calculation precision and so on. 
Moreover, the user can easily change particle masses and form factors
or append new ISs, without changing the code,
since they are recorded in the separate text file,
which is read and interpreted by the program.

\section{Theoretical background}

 \subsection{Kinematics}\label{kin}

  First, we summarize some relations, pertaining to the kinematics
  of the elastic electron-hadron scattering.
  Let $k$ ($k'$) be initial (final) electron momentum,
  and $p$ ($p'$) be the respective momenta of target hadron.
  The transferred momentum is $q=p'-p$,
  the Mandelstam variables are $s=(p+k)^2$, $u=(p-k')^2$,
  and $t = q^2 = -Q^2$.
  The electron mass $m$ is usually neglected,
  and the hadron mass is $M$.

  The kinematics of the elastic scattering of two particles
  can be fully described by two parameters.
  We adopt for this purpose $Q^2$ and $\ve$,
  the virtual photon polarization parameter:
\be
  \ve = \left[ 1 + 2 (1+Q^2/4M^2) \tg^2 (\theta/2) \right]^{-1},
\ee
  where $\theta$ is laboratory scattering angle.
  The parameter $\ve$ can also be expressed as
\be
  \ve = \frac{\nu^2 - Q^2(4M^2+Q^2)}{\nu^2 + Q^2(4M^2+Q^2)},
\ee
  where $\nu = s-u = (p+p')(k+k')$.

 \subsection{Definitions of TPE amplitudes and observables} \label{app:TPEampl}  

 \subsubsection*{Target spin $0^-$}\label{TPE0}
  In general, the elastic scattering amplitude contains one form factor (FF)
  and has the following form
\be
 \CM = - \frac{4\pi\alpha}{q^2} \bar u'\gamma^\mu u \, (p+p')_\mu \, F(q^2,\ve)
\ee
  where $\alpha$ is fine structure constant
  and $u$ ($u'$) are initial (final) electron spinors.
  The one photon exchange (OPE) amplitude has the same form,
  except that the form factor $F$ depends on $q^2$ only.
  Thus we may write
\be
  F(q^2,\ve) = F(q^2) + \delta F(q^2,\ve) + O(\alpha^2)
\ee
  where $\delta F$ is TPE amplitude.
  The quantity output by the program is "normalized amplitude" $\delta F/F$.
  
  The TPE correction to the cross-section is obviously
\be
  \frac{\delta\sigma}{\sigma} = 2 \Re \frac{\delta F}{F}
\ee
   
 \subsubsection*{Target spin $1/2^+$}\label{TPE12}
  The elastic scattering amplitude has the following structure
\be
  \CM = -\frac{4\pi\alpha}{q^2} \bar u'\gamma^\mu u \,
 \bar U' \left[\gamma_\mu F_1(q^2,\ve)
 - [\gamma_\mu,\hat q] \frac{F_2(q^2,\ve)}{4M}
 + (p_\mu+p'_\mu) (\hat k + \hat k') \frac{F_3(q^2,\ve)}{4M^2}
 \right] U
\ee
  (where $U$ and $U'$ are initial and final spinors for the target),
  whereas in OPE approximation $F_3$ is absent,
  and $F_1$, $F_2$ depend on $q^2$ only.

  Instead of $F_1$, $F_2$, $F_3$, it is convenient to introduce
\be
 \begin{array}{ccl}
  \CG_E & = & F_1 - \tau F_2 + \nu F_3/4M^2\\
  \CG_M & = & F_1 + F_2 + \ve \nu F_3/4M^2\\
  \CG_3 & = & \nu F_3/4M^2
 \end{array}
\ee
  In OPE approximation $\CG_E$ and $\CG_M$ become
  usual electric and magnetic FFs, thus
\be
\begin{array}{ccl}
 \CG_E(q^2,\ve) & = & G_E(q^2) + \delta\CG_E(q^2,\ve)\\
 \CG_M(q^2,\ve) & = & G_M(q^2) + \delta\CG_M(q^2,\ve)\\
 \CG_3(q^2,\ve) & = & \delta\CG_3(q^2,\ve)
\end{array}
\ee
  The cross-section is more simply expressed 
  through this set of amplitudes, see below.
  It is common to normalize TPE amplitudes for the electron-proton scattering
  by the proton magnetic FF, thus the program output is
  $\delta\CG_E/G_M$, $\delta\CG_M/G_M$, $\delta\CG_3/G_M$.

  The correction to unpolarized cross-section will be
\be
 \frac{\delta\sigma}{\sigma} = 
 \frac{2}{\ve R^2 + \tau} \Re \left\{ \ve R^2 \frac{\delta\CG_E}{G_E} + \tau \frac{\delta\CG_M}{G_M} \right\}
\ee
where $\tau = Q^2/4M^2$ and $R=G_E/G_M$.
The following relations may prove useful for polarization transfer experiments:
the corrections to measured FF ratio $R_{exp}$,
\be
 \frac{\delta R_{exp}}{R_{exp}} = \Re \left\{ \frac{\delta\CG_E}{G_E}
    - \frac{\delta\CG_M}{G_M}
    - \frac{\ve(1-\ve)}{1+\ve} \frac{\delta\CG_3}{G_M} \right\}
\ee
and to longitudinal component of final proton polarization, $P_l$,
\be
 \frac{\delta P_l}{P_l} = - 2\ve \Re \left\{
   \frac{R^2}{\ve R^2 + \tau}
     \left( \frac{\delta\CG_E}{G_E} - \frac{\delta\CG_M}{G_M} \right)
   + \frac{\ve}{1+\ve} \frac{\delta\CG_3}{G_M}
 \right\}
\ee

  \subsection{Method outline}

In this section we briefly describe the calculation procedure
and the formulae used.

First, the imaginary part of the scattering amplitude can be written
\be \label{uni}
 \Im \CM = \frac{1}{8\pi^2} \sum_{R} \int
    \frac{(4\pi\alpha)^2}{t_1 t_2}
    \bar u' \gamma_\mu (\hat k'' + m) \gamma_\nu u \,
    \sum_{spin} \CM_\mu^{(R)} {\CM_\nu^{(R)}}^* \,
    \theta(p''_0) \delta(p''^2-M_R^2)
    \theta(k''_0) \delta(k''^2) d^4k''
\ee
where $p''=p+k-k''$, $t_1 = (k''-k)^2$, $t_2 = (k''-k)^2$,
and $\CM_\mu^{(R)}$ are amplitudes for the excitation of the
hadronic state $R$ by virtual photon
(i.e. $T\gamma^* \to R$, where $T$ stands for the target).
They contain {\it on-shell} transition FFs and are explicitly
written down in Appendix~\ref{app:ampl}.

The full TPE amplitude is then reconstructed
using analyticity property.
Finally, 
the TPE amplitude, say, $\delta F$, is expressed as
\be
 \delta F(\nu) = \sum_R \left[ \delta F^{(R)}_{box}(\nu) - \delta F^{(R)}_{box}(-\nu)\right],
\ee
where 
\bea
 \delta F^{(R)}_{box}(\nu) &=& \frac{1}{[\nu^2+t(4M^2-t)]^2}
   \sum_{i,j} \int d^4 k''\frac{F_i^{(R)}(t_1)F_j^{(R)}(t_2)}{t_1 t_2} \times \\
 && \times \left[
     \frac{A_{ij}^{(R)}(t_1,t_2,\nu,t)}{(k''^2-m^2)(p''^2-M_R^2)}
   + \frac{A_{k,ij}^{(R)}(t_1,t_2,\nu,t)}{k''^2-m^2}
   + \frac{A_{p,ij}^{(R)}(t_1,t_2,\nu,t)}{p''^2-M_R^2}
   + A_{1,ij}^{(R)}(t_1,t_2,\nu,t)
 \right]\nonumber
\eea
where $F_i^{(R)}$ are FFs for the electromagnetic excitation of the IS $R$.
The quantities $A_{ij}^{(R)}$, $A_{k,ij}^{(R)}$, $A_{p,ij}^{(R)}$
and $A_{1,ij}^{(R)}$ are polynomials in their arguments
and actually depend on the IS $R$ only via its spin-parity.

Parameterizing transition FFs as a sum of poles
\be
  F_i^{(R)}(t) = \sum_a \frac{c_{ia} t}{t-m_a^2},
\ee
as assuming they decrease at $t \to \infty$ fast enough,
the above formula may be rewritten as
\bea
 \delta F^{(R)}_{box} &=& \sum_{ijab} c_{ia} c_{jb}
   \int \frac{d^4 k''}{(t_1-m_a^2)(t_2-m_b^2)} \times \\
  && \times \left\{ 
     \frac{A_{ij}^{(R)}(m_a^2,m_b^2,\nu,t)}{(k''^2-m^2)(p''^2-M_R^2)}
   + \frac{A_{k,ij}^{(R)}(m_a^2,m_b^2,\nu,t)}{k''^2-m^2}
   + \frac{A_{p,ij}^{(R)}(m_a^2,m_b^2,\nu,t)}{p''^2-M_R^2}
   + A_{1,ij}^{(R)}(m_a^2,m_b^2,\nu,t)
 \right\}   \nonumber
\eea
%
%
%
    
\section{Program usage}
 \subsection{General}
   The program is invoked as follows
\begin{verbatim}
   TPEcalc [options] parfile
\end{verbatim}
   All available options are described in detail in Sec.~\ref{Sec:options}.
   Mandatory parameter {\tt parfile} specifies the file, which contains
   the information about the target and ISs, such as masses, spin-parities,
   transition FFs.
   Three predefined parfiles come with the program: {\tt proton.par},
   {\tt neutron.par} and {\tt pion.par},
   for proton, neutron and pion targets respectively
   (see description in Table~\ref{tab:parfiles}).
\begin{table}[b]
  \centering
  \begin{tabular}{|c|c|c|c|}
    \hline
    parfile & IS label & particle  & FFs, Ref. \\
    \hline
    \hline
    \tt proton.par  & \tt proton & $p$            & \cite{Arrington}\\ 
                    & \tt Delta  & $\Delta(1232)$ & \cite{ourDelta} \\    
    \hline
    \tt neutron.par & \tt neutron & $n$     & \cite{Kelly}\\ 
    \hline
    \tt pion.par    & \tt pion & $\pi^+$     & \cite{ourPi} \\
                    & \tt rho  & $\rho$      & \cite{ourPi} \\
                    & \tt b1   & $b_1(1235)$ & \cite{ourPi} \\
    \hline                       
  \end{tabular}  
\caption{}\label{tab:parfiles}
\end{table}

   At first, the program reads and verifies particle descriptions 
   from the parfile. If everything is fine, it prints a short message,
   like this
\begin{verbatim}
   TPEcalc version 1.00
   Current options: -i= -o= -h= -f=a -n=1000 -c=0.000 -m=0 proton.par
   Loaded 2 particle(s):
   (x) proton: M=0.938270, JP=1/2+
   ( ) Delta: M=1.232000, JP=3/2+
\end{verbatim}
   A mark on the left of the IS label indicates that
   the contribution of this IS will be calculated and added to the result.
   The particle masses here (as well as other dimensional parameters
   like $Q^2$ below) are in units of GeV ($\!\GeV^2$).
   
   After that, the program reads $Q^2$ and $\ve$ values,
   either from the {\tt stdin}
   or from the file (if {\tt -i} option is present)
   and calculates corresponding TPE amplitudes.
   The amplitudes are output in the table form either to {\tt stdout}
   or to the file ({\tt -o} option). 
   The table columns are,
   \begin{itemize}
   \item for target spin $0^-$:
   \[
   \begin{array}{ll}
     \Re\frac{\delta F}{F} & \Im\frac{\delta F}{F}
   \end{array}
   \]
   \item for target spin $1/2^+$:
   \[
   \begin{array}{llllll}
      \Re\frac{\delta\CG_E}{G_M} & \Im\frac{\delta\CG_E}{G_M} &
      \Re\frac{\delta\CG_M}{G_M} & \Im\frac{\delta\CG_M}{G_M} &
      \Re\frac{\delta\CG_3}{G_M} & \Im\frac{\delta\CG_3}{G_M}
   \end{array}   
   \]
   \end{itemize}
   The definitions of the amplitudes $\delta F$ and
   $\delta\CG_E$, $\delta\CG_M$, $\delta\CG_3$ are given
   in Sec.~\ref{TPE0}.
   The above-described output format may be overridden
   with the {\tt -f} option.
   
   The elastic contribution (i.e. that of the IS coinciding with the target)
   is well-known to be infra-red divergent.
   If the elastic contribution is included, standard Mo\&Tsai correction
   (\cite{Tsai}, see also \cite{ourPi})
   will be automatically subtracted to cancel the divergence.

 \subsection{Options} \label{Sec:options}
  \subsubsection*{{\tt -i=FILE}, input file}
    If this option is present, program takes $Q^2$ and $\ve$ values
    from the file {\tt FILE}, otherwise from {\tt stdin}.
  \subsubsection*{{\tt -o=FILE}, output file}
    If this option is present, resulting TPE amplitudes are
    written to the file {\tt FILE}, otherwise to {\tt stdout}.  
  \subsubsection*{{\tt -h=IS1,...,ISn}, intermediate states to include}
    This option allows to specify which ISs
    (of those contained in the parfile) will be included in the calculation.
    One should supply comma-separated list of IS labels with no spaces
    between them.
    If some IS is not found in the parfile,
    it is ignored and a warning message is printed.
    In absence of this option, ISs are included or skipped
    according to the defaults in the parfile (see Sec.~\ref{Sec:parfile}).
  \subsubsection*{{\tt -f=[qeaf]}, output format}
    By default, the program prints only calculated TPE amplitudes
    (real and imaginary parts).
    The {\tt -f} option allows to customize the output.
    One should supply a combination of letters
    {\tt q},{\tt e},{\tt a},{\tt f},
    which specify the columns of the output table
    as follows: {\tt q} = $Q^2$, {\tt e} = $\ve$, {\tt a} = TPE amplitudes,
    {\tt f} = elastic FFs of the target.
    E.g. with {\tt -f=qea} the program will print    
   \[
   \begin{array}{llllllll}
      Q^2 &\ve & \Re\frac{\delta\CG_E}{G_M} & \Im\frac{\delta\CG_E}{G_M} &
      \Re\frac{\delta\CG_M}{G_M} & \Im\frac{\delta\CG_M}{G_M} &
      \Re\frac{\delta\CG_3}{G_M} & \Im\frac{\delta\CG_3}{G_M}
   \end{array}
   \]   
   with {\tt -f=qf} no TPE will be calculated and   
   \[
   \begin{array}{lll}
    Q^2 & F_1(Q^2) & F_2(Q^2)
   \end{array}
   \]   
   will be printed (assuming $1/2^+$ target).   
    
  \subsubsection*{{\tt -n=NUMBER}, calculation precision}
    The loop integrals, needed for the calculation,
    are evaluated semianalytically
    (using numerical integration as a final step).
    The precision of the latter can be controlled
    by the {\tt -n} option.
    The {\tt NUMBER} must be positive integer;
    the greater the number, the more precise
    (but also more time-consuming) the calculation will be.
    The default is {\tt -n=1000}.
    
  \subsubsection*{{\tt -c=NUMBER}, form factor cut-off}  
    For the program to work correctly,
    transition FFs should decrease fast enough as $Q^2 \to \infty$.
    If this is not the case, the program will print error message and exit.
    To overcome this problem, one may use {\tt -c} option.
    The FFs specified in the parfile will be damped according to
    \be
     F(Q^2) \to F(Q^2) \frac{\Lambda^2}{Q^2+\Lambda^2}
    \ee
    where $\Lambda$ (in units of GeV) is the number,
    specified after {\tt -c}.
    
  \subsubsection*{{\tt -m=NUMBER}, electron mass}
    This option is experimental.
    It sets non-zero electron mass (in units of GeV).
    In the current version, the only place,
    where the electron mass is present,
    is propagator denominators.
    In any case it is neglected in the numerators.
   
 \subsection{Parfile format} \label{Sec:parfile}
   The parfile contains the information about
   the target and intermediate states, involved in the calculation.
   It is a plain text file, and the user may edit it or create new parfiles
   adhering to the format, described below.
      
   A line, which begins with a \% sign, is a comment (ignored by the program).
   
   The very first line in the file contains version number,
   which must match the program version.
   This is necessary because the parfile format may change in the future versions.
   Please do not modify this line.  

   The rest of the file consists of one or more particle descriptions,
   according to the following format.
   The first particle in the file is the target, the others are inelastic ISs.
   Here is an example:
\begin{verbatim}
   pion : 1
   M=0.13957 JP=0-
   % this line is a comment
   0       1.
   0.7194 -1.
   
   rho : 0
   M=0.7758 JP=1-
   0        0.725333333
   0.78259 -0.725333333
\end{verbatim}
   Each particle description begins with a label, followed by a colon.
   The label is used solely to identify the particle, and may be any word,
   convenient to user.
   It should consist of alphanumeric characters
   and must not contain spaces or commas.
   The maximum label length is 15 characters. 
   A digit following the colon is the "inclusion flag".
   If it is non-zero, then the contribution of this IS is calculated
   and added to the resulting TPE amplitude.
   Otherwise, the IS is skipped.
   This behaviour may be overridden with the {\tt -h} option.
   
   Next line specifies mass (in units of GeV)
   and spin-parity of the particle.
   The mass can be any positive number, supported spin-parities
   can be found in Appendix~\ref{app:ampl}.
   
   The subsequent lines contain a matrix of the transition FFs
   (for inelastic IS) or elastic FFs (for elastic IS),
   terminated by a blank line or end-of-file.
   The matrix has the form
   \be
     \begin{array}{ll@{\quad...\quad}l}
       m_1 & c_{11} & c_{k1} \\
       m_2 & c_{12} & c_{k2} \\
       ... & ... & ... \\
       m_n & c_{1n} & c_{kn}
     \end{array}
   \ee
   and defines the transition FFs via
   \be
    F_i(Q^2) = \sum_{a=1}^n \frac{c_{ia} Q^2}{Q^2+m_a^2}
   \ee   
   Thus the number of columns in the matrix is (number of transition FFs) + 1.
   The first row should always contain zero mass, $m_1=0$, and 
   the coefficients $c_{i1} = F_i(0)$.   
   The corresponding transition amplitudes in terms of FFs $F_i$
   for each supported spin-parity
   are written down in Appendix~\ref{app:ampl}.  
 
 \subsection{Limitations}  
   
   The input parameters must satisfy $Q^2>0$, $0<\ve<1$.
   In the low-$\ve$ region, the accuracy drops seriously as $\ve\to 0$,
   because the calculation method involves a subtraction at $\ve=0$.
   Setting higher precision with {\tt -n} option may help.
   Though $\ve=0$ is legal kinematics (strictly backward scattering),
   the calculation at $\ve=0$ is impossible
   due to the above property.
   
   With the inelastic ISs, the accuracy also drops
   when the total c.m. energy squared $s$
   is near the resonance, $s\sim M_R^2$.
   For $m=0$ TPE amplitudes diverge as $\ln |s-M_R^2|$ near that point.   
   This is an artifact, arising from neglecting IS width.
   Actually, if one takes finite width into account, the $s$ dependence
   become "smeared" and the divergence should disappear.
   However, this is not possible in the present version of the program.


\section{Compiling the program}

We have successfully compiled {\tt TPEcalc} with GNU {\tt gcc} ({\tt g++}),
under both Linux and Windows,
but compiling with any other C++ compiler
should constitute no problem.
First, unpack {\tt TPEcalc.zip} into an empty directory.
You should have 17 source files (of which 6 in the subdirectory {\tt A/}),
3 parfiles ({\tt proton.par}, {\tt neutron.par}, {\tt pion.par}),
and {\tt README} and {\tt makefile}.
Then run {\tt make} to build the {\tt TPEcalc} executable.
If you use some other compiler (not {\tt gcc}),
you should adjust two first lines of the makefile:
\begin{verbatim}
 CXX = <your C++ compiler>
 options = <compiler options>
\end{verbatim}
It is good idea to set maximum optimization,
to ensure best performance.

The parfiles should be in the current directory when running {\tt TPEcalc},
otherwise you must provide full path to them.
E.g. under Linux you might want to place {\tt TPEcalc} and parfiles into
{\tt \$HOME/bin}, and then invoke the program like this
\begin{verbatim}
  TPEcalc [options] ~/bin/proton.par
\end{verbatim}

\appendix

\section{Structure of the transition amplitudes} \label{app:ampl}

This section contains the formulae for electromagnetic transition amplitudes 
between supported targets and ISs:
\be
T(p-q) + \gamma^*(q) \to R(p),
\ee
where $T$ is target and $R$ is IS, and particles' momenta are shown in parentheses.
Below, $M_R$ is the mass of the IS, $M$ is target mass.
In each case, the program assumes that the parfile contains
corresponding form factors, $F_1$, $F_2$, etc.

\subsection{Target spin $0^-$}

\begin{itemize}
\item
For the elastic IS:
\be
 \CM_\mu = (2p_\mu-q_\mu) \, F_1
\ee
\item
For pseudovector IS ($J^P=1^-$), e.g. $\rho$ meson:
\be
 \CM_\mu = \frac{2M_R}{M_R^2-M^2}
  \ve^{\mu\nu\sigma\tau} q_\nu p_\sigma v_\tau^* \, F_1
\ee
Here and in the next equation $v$ is spin-1 wavefunction of the IS,
satisfying $p_\alpha v^\alpha = 0$ and $v_\alpha v^\alpha = -1$.
\item
For vector IS ($J^P=1^+$), e.g. $b_1$ meson:
\be
 \CM_\mu = \frac{2M_R}{M_R^2-M^2} 
  (g^{\mu\alpha} q^\nu - g^{\mu\nu} q^\alpha)
       ( p_\alpha F_1 - q_\alpha F_2 ) v_\nu^*
\ee
\end{itemize}

\subsection{Target spin $1/2^+$}
In this subseciton, $U$ is target's bispinor amplitude.
\begin{itemize}
\item
For the elastic IS:
\be
  \CM_\mu = \bar V \left\{
     \gamma_\mu F_1 - [\gamma_\mu,\hat q] \frac{F_2}{4M}
     \right\} U
\ee
i.e. $F_1$ and $F_2$ are usual Dirac and Pauli FFs.
Note that TPE amplitudes will be normalized by $G_M = F_1 + F_2$.
\item
For the IS spin-parity $1/2^\pm$:
\be
  \CM_\mu = \bar V \gamma_5^{\frac{1-P}{2}} \left\{ 
     \left( \gamma_\mu - \frac{q_\mu \hat q}{q^2} \right) F_1
        - [\gamma_\mu,\hat q] \frac{F_2}{4M}
  \right\} U
\ee
where $P=\pm 1$ is parity of the IS, and $V$ is its bispinor amplitude.
\item
For the IS spin-parity $3/2^\pm$ (e.g. Delta resonance):
\be
  \CM_\mu = \frac{1}{2 M^2}
  ( g^{\mu\alpha} q^\nu - g^{\mu\nu} q^\alpha ) \,
    \bar V_\alpha \gamma_5^{\frac{1+P}{2}}  \left\{
        (\hat p \gamma_\nu - p_\nu ) F_1 + p_\nu F_2 - q_\nu F_3
    \right\} U
\ee
where $V_\alpha$ is Rarita-Schwinger wavefunction, normalized according to
$\bar V_\alpha V^\alpha = -2M_R$. Here $F_1$ corresponds to magnetic,
$F_2$ --- to electric and $F_3$ --- to Coulomb form factor.
\end{itemize}


\end{document}